\input{aipcheck}

\documentclass[
    ,final            
  ]
  {aipproc}

\layoutstyle{6x9}

\def\kms{\relax \ifmmode {\,\rm km\,s}^{-1}\else \,km\,s$^{-1}$\fi}
\def\nii{[N {\sc ii}]}
\def\sii{[S {\sc ii}]}
\def\oii{[O {\sc ii}]}
\def\oiii{[O {\sc iii}]}
\def\ha{H$\alpha$}

\begin{document}

\title{Microstructures of Planetary Nebulae with Large Telescopes}

\classification{01.30.Cc}
\keywords      {ISM: jets and outflows - Planetary nebulae: individual: K~4-47 and 
NGC~7009}

\author{Denise R. Gon\c calves}{
  address={Insituto de Astrof\'\i sica de Canarias, Tenerife, Spain  \\
Instituto de Astronomia, Geof\'\i sica e Ci\^encias Atmosf\'ericas - USP, S\~ao
Paulo, Brazil}
}

\begin{abstract}
Planetary nebulae (PNe) are known to possess a variety of	 
small-scale structures that are usually in a lower ionization state 
than the main body of the nebulae.  The morphological and kinematic 
properties of these low-ionization structures (LISs) vary from type	  
to type in the sense that LISs can appear in the form of pairs of	  
knots, filaments, jets, and isolated features moving with velocities	  
that either do not differ substantially from that of the ambient nebula,	  
or instead move supersonically through the environment. The high-velocity jets  
and pairs of knots, also known as FLIERs, are likely to be shock-excited. 
So far, most of the FLIERs analyzed with ground-based small and medium telescopes,	  
as well as with the HST, do not show the expected shock-excited features ---either 
the bow-shock geometry or the shock excited emission lines.	  
In this talk we discuss the crucial problem of the excitation mechanisms	  
of FLIERs ---through the comparison of jets and knots of NGC 7009 and	  
K 4-47--- and what might be the contribution of large telescopes.
\end{abstract}

\maketitle

\section{Macro- and Microstructures of PNe}

The main, macro, structures of PNe are the elliptical, bipolar, or 
point-symmetric rims and shells, as well the haloes that dominate 
the optical \oiii\ and \ha\ line emission of  planetary nebulae 
(see top panel of Figure~1). 
Notwithstanding important revisions, the mechanism responsable 
for the rim formation has been known for almost 30 years as the
interacting stellar wind, ISW, models (Kwok, Purton \& 
FitzGerald 1978). Now we believe that, in addition to the interplay  
between the slow AGB and the fast post-AGB winds, 
magnetic fields and rotation ---probably due to a disk within a 
binary system--- should also play a role in the formation of the 
main structures of the PNe (for a review, see Balick \& Frank 2002). 
The formation of the shells, external to the rims, has been ascribed 
to the action of the photoionization front on the AGB matter not yet 
reached by the shock produced by the fast wind (e.g., Sch\"onberner 2002).  
The outer halo in its turn is interpreted by Corradi et al.\ (2003) as 
being composed by matter ejected during the AGB phase, its outer edge 
marking the signature of the last AGB thermal pulse.
 
\begin{figure}
  \includegraphics[height=.35\textheight]{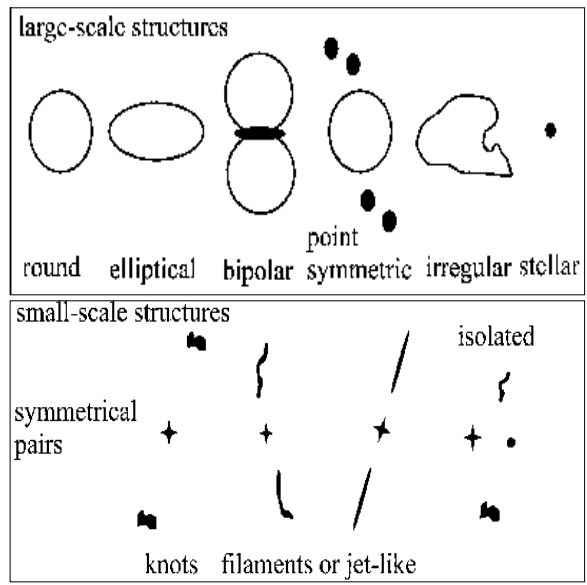}
  \caption{Morphological classes of PNe. Top: The main components. 
  Bottom: The microstructures, LISs.}
\end{figure}

The microstructures of PNe, on the other hand, are  far less understood 
theoretically. That they are much more prominent in the lower 
ionization optical emission lines (\nii, \sii, \oii), appear on smaller 
scales than the main structures, and are usually different of the main bodies 
in terms of morphology and kinematics, gives us some guidelines in attempting 
to unveil their origin. A number of papers have recently been published on the subject, 
in which some aspects of their origin are treated. From  observations, we 
know that:  LISs appear as pairs of jets, knots, filaments, and jetlike features, 
or as isolated systems (bottom panel of Figure~1); 
 LISs sometimes expand with the rim, shells, or haloes in which 
they are embedded, but sometimes they are much faster than the main PNe 
components;
they are spread indistinctly within all the morphological classes 
of PNe;
in general, they do not have an important density contrast with 
respect to the main bodies; and 
most LIS systems studied up to now are mainly photoionized.

With regard to the comparison of LIS properties with  theoretical predictions, 
some of their characteristics seem hard to explain (see Balick \& Dwarkadas 
1998; Gon\c calves et al.\ 2001, 2003; Balick \& Frank 2002). However, the 
origin of part of these systems (from their morphology and kinematics) can 
be reasonably understood via ISW models, in single stars or binaries, 
with or without magnetic fields, precession, and wobbling. 

\section{Looking for Shock Excitation in LIS}

Because the high-velocity pairs of knots and jets are 
highly supersonic, it is expected that their optical emission line 
spectrum show line ratios characteristic of shock-excited emission. 

Figure~2 and 3 show three diagnostic diagrams for K~4-47 and NGC~7009,  
in which the excitation of the different regions of these PNe can be checked. 
From the figures, the high-velocity pair (Knot1--Knot2) of LISs in K~4-47 is mainly 
shock-excited, while its core is locate out of the shock regime zone, much 
closer to the loci of the photoionized structures. In contrast, NGC~7009 
has three pairs of highly supersonic LISs ---the knots K1--K4, K2--K3, and the jets 
J1--J2--- in addition to its rim (R1--R2). In this case, none of the features, 
either LIS or rim, fits the zone of the shock-excited emission. 

As in K~4-47, a few other PNe were found with shock-excited 
features: M~1-16 (Schwarz 1992, Huggins et al.\ 2000); and M 2-48 
(L\'opez-Mart\'\i n et al.\ 2002). What these three PNe have in common is 
that they are highly collimated bipolar PNe; have high-velocity structures 
(100 up to 300~\kms); and share properties with young PNe. NGC~7009, on the 
other hand, seems to be more evolved than K~4-47, M~1-16, and M~2-4.

\begin{figure}
  \includegraphics[height=.4\textheight]{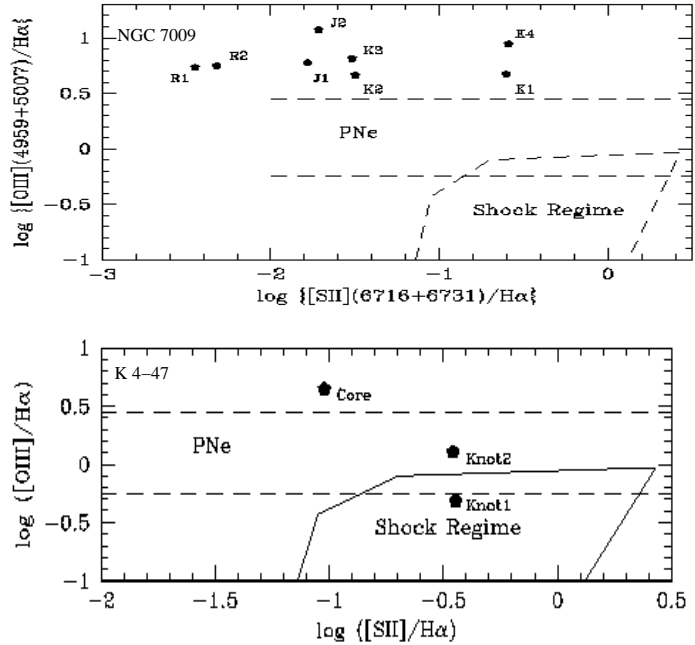}
  \caption{The \oiii/\ha\ {\it vs.} \sii/\ha\ diagram showing the excitation of the 
           eight regions of NGC~7009 (top) and three of K~4-47 (bottom), from  2.5 m INT long-slit 
	   medium-resolution spectra (see Gon\c calves et al.\ 2003, 2004).}
\end{figure}

\begin{figure}
  \includegraphics[height=.4\textheight]{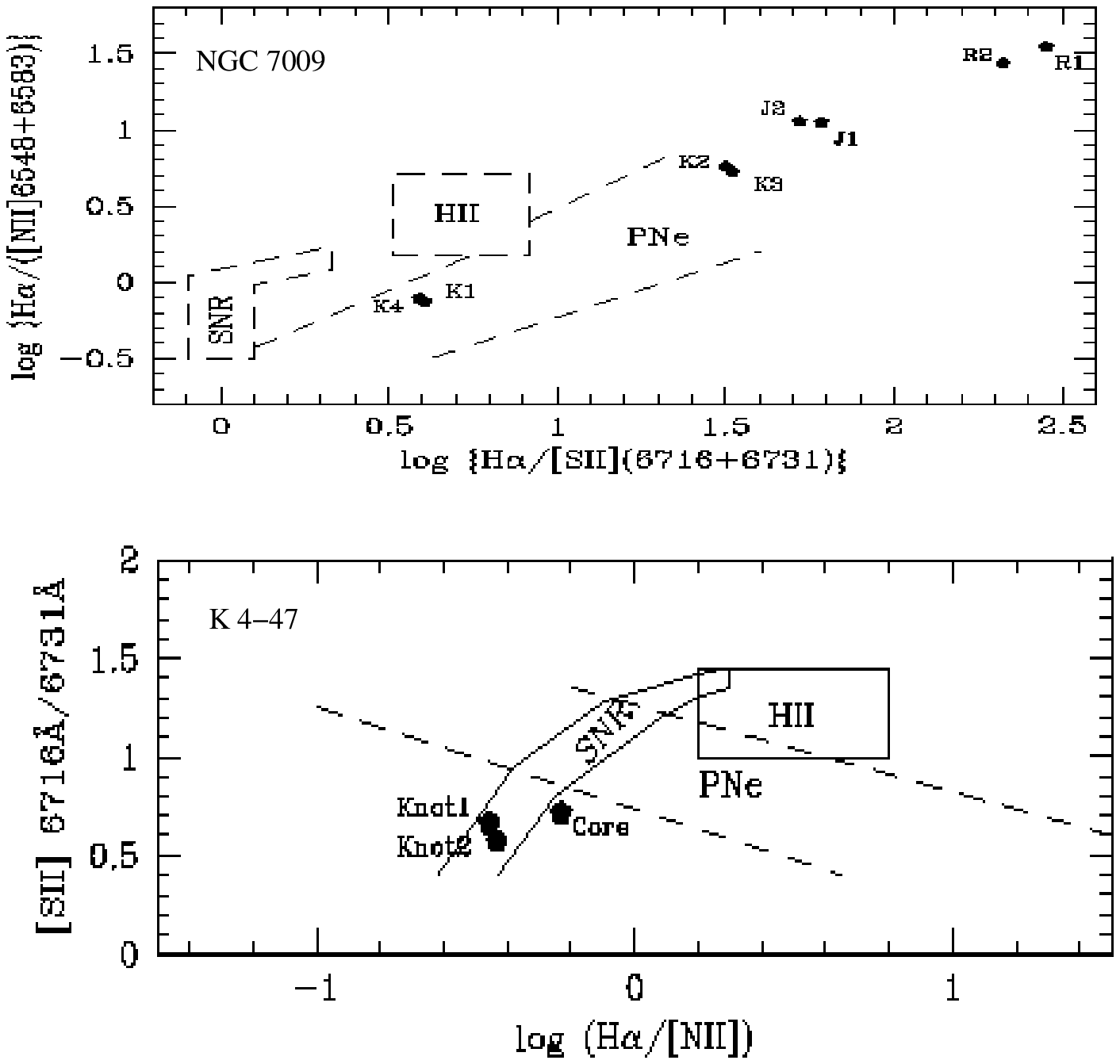}
  \caption{As in Fig. 2, but in the \ha/\nii\ {\it vs.} \ha/\sii\ and 
  \sii(6716\AA)/\sii(6731\AA) {\it vs.} \ha/\nii\ diagrams.}
\end{figure}

\section{What about ``the largest telescopes''?}

The fact that the expected shock excitation of jets and other high-velocity 
LISs is not usually observed ---in evolved PNe--- could mean that jets/knots 
are relaxed systems in the sense that shock excitation  
is no longer present because LISs were already reached 
by the energetic photons of the post-AGB central star, and/or affected by 
local instabilities (Dopita 1997; Miranda et al.\ 2000; Soker \& Reveg 1998). 
This seems to be the case for NGC~7009. In fact the PN NGC~7662  
was observed with the HST (WFPC2 and STIS) and no shock-excited emission was 
found to be associated with its FLIERS (Perinotto et al.\ 2004). Therefore, it is clear 
that either shocks are not present in evolved PNe (not even associated with the highly 
supersonic LISs) or the `the thickness of the shocked layer in LIS is too small to 
be resolved with HST,' as suggested by the latter authors.  

Two ways of further investigating this issue are therefore: i) to determine 
via simulations 
 the size of the shock zone (or the thickness of the working surface 
associated with the shock) of LISs in evolved PNe, and ii) observe these structures with 
the largest telescopes, which would give a spatial resolution better than with the HST.

\begin{theacknowledgments}
I thank the Organizing Committee for the wonderful conference and acknowledge partial support 
from the Spanish Ministry of Science 
and Technology (AYA 2001-1646).
\end{theacknowledgments}


\end{document}